\documentclass[aps,floatfix,superscriptaddress,showpacs,twocolumn]{revtex4}
\usepackage{graphicx}
\usepackage{array}
\usepackage{amsthm}
\usepackage{amsmath}
\usepackage{amssymb}

\begin{document}

\title{Fisher information manifestation of dynamical stability and
transition to self-trapping for Bose-Einstein condensates}
\author{Yixiao Huang}
    \affiliation{Zhejiang Institute of Modern Physics, Department of Physics, Zhejiang
    University, Hangzhou 310027, China}

\author{Wei Zhong}
    \affiliation{Zhejiang Institute of Modern Physics, Department of Physics, Zhejiang
    University, Hangzhou 310027, China}

\author{Zhe Sun}
    \affiliation{Department of Physics, Hangzhou Normal University, Hangzhou 310036, China}

\author{Xiaoguang Wang}
    \email {xgwang@zimp.zju.edu.cn}
    \affiliation{Zhejiang Institute of Modern Physics, Department of Physics, Zhejiang
    University, Hangzhou 310027, China}

\begin{abstract}
    We investigate dynamical stability and self-trapping for
    Bose-Einstein condensates in a symmetric double well.
    The relation between the quantum Fisher information and
    the stability of the fixed point is studied.
    We find that the quantum Fisher information displays a sharp
    transition as the fixed point evolving from stable to unstable regime.
    Moreover, the transition from Josephson oscillation to self-trapping
    is accompanied by an abrupt change of the quantum Fisher information.
\end{abstract}

\maketitle

\section{Introduction}
Quantum Fisher information (QFI) which characterizes the sensitivity
of the state with respect to changes of the parameter,
is a key concept in parameter estimation theory~\cite{R. A. Fisher,V. Giovannetti}.
It has important applications in quantum technology
such as quantum frequency standards~\cite{J. J. Bollinger,S. F. Huelga},
measurement of gravity accelerations~\cite{A. Peters},
and clock synchronization~\cite{R. Jozsa}.
Recently, the QFI was found to be able to detect entanglement and quantum
phase transitions in many-body systems \cite{L. Pezze,Shi-Jian Gu,C. Brukner,S. Y. Cho,Yan Chen}.
In the Lipkin-Meskhov-Glick model,
the QFI can be used to characterize the ground state which displays a second-order
quantum phase transition~\cite{J. Ma}.
The critical point of a transverse Ising chain can also be estimated by QFI~\cite%
{C. Invernizzi,Z. Sun}.

We use QFI to study the dynamical stability and the transition from
Josephson oscillation (JO) to self-trapping (ST) for Bose-Einstein
condensates (BECs) in a double well potential. This transition is an
interesting finding in BECs~\cite{M. H. Anderson,K. B. Davis,C. C.
Bradley,G. J. Milburn,A. Smerzi,S. Raghavan}. By changing the atomic
interaction, the Josephson oscillation may be blocked, and the atoms
of a BEC in a symmetric double-well potential may show a highly
asymmetric distribution between two wells. This phenomenon has been
observed in the experiment~\cite{M. Albiez}.

Over the past few years, people found the transition is strongly
related to the fixed points of the underlying classical dynamics.
This is because quantum system can be mapped into a classical
Hamiltonian which gives rise to fixed point solutions~\cite{K. W.
Mahmud}. Recently, some researchers have explored that quantum
entanglement which is closely related to QFI can manifest the
transition~\cite{A. P. Hines,L. B. Fu,J. Vidal}. People also found
quantum entanglement has relation to the classical fixed-point
bifurcation$-$a loss of stability and the emergence of new fixed
points \cite{R. Wu,S. Schneider,A.P. Hines,X.W. Hou}. However,
effects of the stabilities of the fixed points on quantum
information flow is still lacking. It is thus certainly important to
explore what features of the dynamical stabilities are manifested by
the QFI.

We study the Hamiltonian of the system with the approach of spin
coherent state (SCS) associated to SU(2) group. We localize all the
fixed points of the model in the phase space and discuss their
stabilities. We also analyze the transition from JO to ST regime. In
particular, we focus on the dynamics of the QFI for different
interaction regions, such as ST and JO, stable and unstable regimes.
We find that QFI can clearly demonstrate the dynamical stability and
the transition from JO to ST.

Our paper is organized as follows.
In Sec.\thinspace II,
we discuss the parameter estimation, QFI, and introduce the maximal mean QFI.
In Sec.\thinspace III,
we study the stabilities of the fixed points, and discuss the transition from JO to ST.
Then in Sec.\thinspace IV,
we investigate the maximal mean QFI for two different initial states and reveal
QFI manifestation of the stability of the fixed point and the transition.
Finally, a summary is provided in Sec.\thinspace V.

\section{Quantum Fisher information}

In this section, we discuss the QFI and the maximal mean QFI.
Generally, for an input state $\rho _{\text{in}}$ under a linear rotation by an angle $\varphi $,
the output state can be written as $\rho_{\varphi}%
=e^{i\varphi J_{\vec{n}}}\rho _{\text{in}}e^{-i\varphi J_{\vec{n}}}$.
According to the Quantum Cramer-Rao theorem,
the phase sensitivity $\varphi $ has a lower bound limit~%
\cite{C. W. Helstrom,A. S. Holevo}%
\begin{equation}
    \Delta \hat{\varphi}\geqslant \Delta \varphi _{\text{QCR}}=\frac{1}{\sqrt{%
    vF(\rho _{\text{in}},J_{\vec{n}})}}\text{,}  \label{CR}
\end{equation}%
where $\hat{\varphi}$ is an unbiased estimator (i.e., $\hat{\varphi}$ = $\varphi $),
$v$ is the number of experiments,
and $F(\rho _{\text{in}},J_{\vec{n}})$
denote the QFI which is defined as~\cite{C. W. Helstrom,A. S. Holevo}
\begin{equation}
    F(\rho _{\text{in}},J_{\vec{n}})=\text{Tr(}\rho _{\varphi }L_{\varphi }^{2}%
    \text{)}  \label{fisher_information}.
\end{equation}%
Here, $L_{\varphi }$ is the so-called symmetric logarithmic derivative
determined by the following equation%
\begin{equation}
    \frac{\partial \rho_\varphi}{\partial \varphi }=\frac{1}{2}\left( \rho _{\varphi
    }L_{\varphi }+L_{\varphi }\rho _{\varphi }\right) \text{.}
    \label{logarithmic}
\end{equation}%
In Eq.\thinspace (\ref{CR}), one notices that,
besides increasing the number of experimental times $v$,
we can improve the estimation precision $\Delta \hat{\varphi}$ by
choosing a proper state for a given $J_{\vec{n}}$,
which maximize the value of the QFI.

Now, we consider the maximal QFI for a given state.
Based on Eq.\thinspace ($\ref{logarithmic}$),
the expression of Eq.\thinspace (\ref{fisher_information}) is explicitly derived as%
\begin{equation}
    F(\rho _{\text{in}},J_{\vec{n}})=2\sum_{i\neq j}\frac{\left(
    p_{i}-p_{j}\right) ^{2}}{p_{i}+p_{j}}\left \vert \left \langle i\right \vert
    J_{\vec{n}}\left \vert j\right \rangle \right \vert ^{2}\text{,}
    \label{QFI}
\end{equation}%
where $ \left\{\left \vert i\right \rangle\right\}$ are the eigenstates of $\rho_{\varphi}$ with eigenvalues
$\left\{p_{i} \right \}.$
Then the Eq.\thinspace (\ref{QFI}) can be compactly rewritten as%
\begin{equation}
    F(\rho _{\text{in}},J_{\vec{n}})=\vec{n}\mathbf{C}\vec{n}^{T}\text{,}
\end{equation}%
where the matrix element for the symmetric matrix $\mathbf{C}$ is%
\begin{equation}
    \mathbf{C}_{kl}=\sum_{i\neq j}\frac{\left( p_{i}-p_{j}\right) ^{2}}{%
    p_{i}+p_{j}}\left[ \left \langle i\right \vert J_{k}\left \vert j\right
    \rangle \left \langle j\right \vert J_{l}\left \vert i\right \rangle +\left
    \langle i\right \vert J_{l}\left \vert j\right \rangle \left \langle j\right
    \vert J_{k}\left \vert i\right \rangle \right] \text{.}
\end{equation}%
It can be seen that the rotation along the $\vec{n}$ direction affects the
sensitivity of the state $\rho $. For a pure state, the QFI can be expressed
as $F(\rho _{\text{in}},J_{\vec{n}})=4(\Delta J_{\vec{n}})^{2}$~\cite{S. L.
Braunstein}. To obtain the maximal QFI, we rewritten the variance as%
\begin{equation}
    (\Delta J_{\vec{n}})^{2}=\vec{n}O(O^{T}\mathbf{C}O)O^{T}\vec{n}^{T}=\vec{n}%
    ^{\prime }\mathbf{C}_{d}\vec{n}^{\prime ^{T}}\text{,}
\end{equation}%
where $O$ is an orthogonal matrix, $\vec{n}^{\prime }$ is a new direction
defined as $\vec{n}^{\prime }=\vec{n}O$, and $\mathbf{C}_{d}$ is the
diagonal form of $\mathbf{C}$,%
\begin{equation}
    \mathbf{C}_{d}=O^{T}\mathbf{C}O=\text{diag}\{ \lambda _{1},\lambda
    _{2},\lambda _{3}\} \text{,}
\end{equation}%
where the $\lambda _{i}$'s are the eigenvalues of $\mathbf{C}$. Now the
maximal variance reads%
\begin{equation}
    \max (\Delta J_{\vec{n}})^{2}=\max \left[ \lambda _{1}(n_{1}^{\prime
    })^{2}+\lambda _{2}(n_{2}^{\prime })^{2}+\lambda _{3}(n_{3}^{\prime })^{2}%
    \right] \text{.}
\end{equation}%
In the above equation, the rotated direction is normalized and satisfies the
condition $n_{1}^{\prime 2}+n_{2}^{\prime 2}+n_{3}^{\prime 2}=1$. If we
set $\lambda _{\text{max}}=\lambda _{1}$ as the maximal one of the eigenvalues, then $\vec{n}%
^{\prime }=(1,0,0)$, and the original direction $\vec{n}=\vec{n}^{\prime
}O^{T}$.

Now we get the maximal QFI%
\begin{equation}
    F_{\max }=4\lambda _{\max }\text{.}
\end{equation}%
For simplicity, we study the maximal mean QFI%
\begin{equation}
    \bar{F}_{\max }=\frac{F_{\max }}{N}\text{,}
\end{equation}%
where $N$ is the number of atoms.
\begin{table*}
\caption{Fix points and stable regimes}

\centering{}
\begin{tabular}{|c|c|c|c|c|}
\hline
Parameters  & a  & b  & c & d
\tabularnewline
\hline
\parbox[c]{2.4cm}{%
$\Omega >2\kappa _{r}$%
}&%
\parbox[c]{2.9cm}{%
 \vspace{1.3mm}
$\theta =\pi /2,$ $\phi =0$ \vspace{1.3mm}%
\\ stable \vspace{1.3mm} } & %
\parbox[c]{2.9cm}{%
 \vspace{1.3mm}
$\theta =\pi /2,$ $\phi =\pi$ \vspace{1.3mm}%
\\ stable \vspace{1.3mm}}  & %
\parbox[c]{4.5cm}{%
 N/A \vspace{1.3mm}
}& %
\parbox[c]{4.5cm}{%
N/A \vspace{1.3mm}
}
\tabularnewline
\hline

\parbox[c]{2.4cm}{%
\vspace{1.3mm}
 $\Omega <2\kappa _{r}$
}& %
\parbox[c]{2.9cm}{%
\vspace{1.3mm}
$\theta =\pi /2,$ $\phi =0$ \vspace{1.3mm}
\\ unstable \vspace{1.3mm} }& %
\parbox[c]{2.9cm}{%
\vspace{1.3mm}
$\theta =\pi /2,$ $\phi =\pi$ \vspace{1.3mm}%
\\ stable \vspace{1.3mm} }& %
\parbox[c]{4.5cm}{%
\vspace{1.3mm}
$\theta=\text{arcsin}[\Omega/(2\kappa_{r})],$ $\phi =0$\vspace{1.3mm}%
\\ stable  \vspace{1.3mm} }
& %
\parbox[c]{4.5cm}{%
\vspace{1.3mm}
$\theta=\pi-\text{arcsin}[\Omega/(2\kappa_{r})],$ $\phi =0$\vspace{1.3mm}%
\\ stable \vspace{1.3mm} }
\tabularnewline
 \hline
\end{tabular}
\end{table*}
\section{Model and a classical analogue}

System of BECs trapped in a symmetric double well have been well studied in
theories and experiments~\cite{M. Grifoni,O. Zobay,F. Kh. Abdullaev,F. Meier}.
Due to the interaction between two wells, the system presents JO and nonlinear ST phenomena~\cite{G. J. Milburn,A. Smerzi,S. Raghavan}.
For the two weakly coupled BECs system,
the Hamiltonian can be described as~\cite{J. I. Cirac,M. J. Steel,L. M. Kuang}%
\begin{equation}
    H=\Omega J_{x}+2\kappa J_{z}^{2},
\end{equation}%
where the angular momentum operators are defined in terms of the creation and
annihilation boson operators $\hat{a}_{1,2}^{\dagger }$, $\hat{a}_{1,2}$ as%
\begin{eqnarray}
    J_{x} &=&\frac{\hat{a}_{1}^{\dagger }\hat{a}_{2}+\hat{a}_{2}^{\dagger }\hat{a%
    }_{1}}{2}, \\
    J_{y} &=&\frac{\hat{a}_{1}^{\dagger }\hat{a}_{2}-a_{2}^{\dagger }\hat{a}_{1}%
    }{2i}, \\
    J_{z} &=&\frac{\hat{a}_{1}^{\dagger }\hat{a}_{1}-\hat{a}_{2}^{\dagger }\hat{a%
    }_{2}}{2},
\end{eqnarray}%
which obey the SU$(2)$ Lie algebra.
The parameter $\Omega $ describes the coupling between two wells, and
$\kappa $ denotes the effective interaction of atoms.
In present work, we focus on the interaction strength $\kappa>0 $ and $\Omega>0 $.
In the system, the total particle number
$N=\hat{a}_{1}^{\dagger }\hat{a}_{1}+\hat{a}_{2}^{\dagger }\hat{a}_{2}$
is a conserved quantity.

To obtain the classical dynamics approach, we use a generalized SCS
as an initial state, which is defined formally as~\cite{J. M. Radcliffe,F. T. Arecchi,T. F. Viscondi}%
\begin{eqnarray}
    \left \vert \theta ,\phi \right \rangle &=&e^{-i\theta (J_{x}\sin \phi
    -J_{y}\cos \phi )}\left \vert j,-j\right \rangle  \notag \\
    &=&\sum \limits_{m=-j}^{j}\left( _{j+m}^{\text{ }2j}\right) ^{1/2}\frac{\tau
    ^{m+j}}{(1+\left \vert \tau \right \vert ^{2})^{j}}\left \vert j,m\right
    \rangle \text{,}  \label{CSS}
\end{eqnarray}%
where $j$ is the angular momentum quantum number, $j=N/2$, $\tau
=e^{-i\phi }\tan \frac{\theta }{2}$, and $\left( _{j+m}^{\text{ }2j}\right) $
is the binomial coefficients.
Under this SCS, the expectation values of the angular momenta are given by
\begin{equation}
    \left \langle \theta ,\phi \right \vert \vec{J}\left \vert \theta ,\phi
    \right \rangle =\frac{N}{2}\left( \sin \theta \cos \phi ,\sin \theta \sin
    \phi ,-\cos \theta \right) \text{.}
\end{equation}%
Meanwhile, we obtain the rescaled Hamiltonian (with constant terms dropped)%
\begin{eqnarray}
    \mathcal{H} &\mathcal{\equiv }&\left \langle \theta ,\phi \right \vert \hat{H%
    }\left \vert \theta ,\phi \right \rangle /j  \notag \\
    &=&\Omega \sin \theta \cos \phi +\kappa _{r}\cos ^{2}\theta \text{,}
\end{eqnarray}%
where $\kappa _{r}=(N-1)\kappa $.
With the help of path integral in the representation of SCS,
we get the Lagrangian $\mathcal{L} $ for the system~\cite{H. Kuratsuji}%
\begin{equation}
    \mathcal{L=}N/2[\hbar (1-\cos \theta )\dot{\phi}-\mathcal{H}],
    \label{Lagrangian}
\end{equation}%
which is associated with canonical coordinate $\phi $ and
canonical momentum $p_{\phi }=\hbar(1-\cos \theta) $ \cite{Yi Zhou}.
For simplicity, we use $p_{\phi }=-\hbar \cos \theta $ as the canonical momentum.
By setting $\hbar =1$, the Hamiltonian can be rewritten
\begin{eqnarray}
    \mathcal{H}&=&\Omega \sqrt{(1-p_{\phi }^{2})} \cos \phi +\kappa _{r}p_{\phi }^{2} \text{.}
\end{eqnarray}%
Then we obtain the following canonical Hamiltonian's equations
of motion for $p_{\phi }$ and $\phi $ in the phase space%
\begin{eqnarray}
    \dot{p}_{\phi } &=&\Omega \sqrt{(1-p_{\phi }^{2})}\sin \phi \text{,}
    \label{dynamic1} \\
    \dot{\phi} &=&2\kappa _{r}p_{\phi }-\frac{\Omega p_{\phi }\cos \phi }{\sqrt{%
    1-p_{\phi }^{2}}}.  \label{dynamic2}
\end{eqnarray}%
It is noted that the motion governed by the above equations are similar to that described by the mean field approximation~%
\cite{G. J. Milburn,A. Smerzi}.

The stationary state solution of equations (\ref{dynamic1}) and (\ref%
{dynamic2}) can be obtained by assuming $\dot{p}_{\phi }=\dot{\phi}=0$.
For the case of $%
\dot{p}_{\phi }=0$,
we get $p_{\phi }=1$, $%
\phi =0$, and $\phi =\pi $.
For $\dot{\phi}=0$, we obtain
$p_{\phi }=0$, $p_{\phi }=$ $\sqrt{1-(
\frac{\Omega }{2\kappa _{r}}\cos \phi ) ^{2}}$, $p_{\phi }=$ $-\sqrt{%
1-( \frac{\Omega }{2\kappa _{r}}\cos \phi ) ^{2}}$.
Under these conditions, we obtain several fixed points and list them in Table I.
It clearly shows that for stronger interaction,
i.e. $\Omega >2\kappa _{r}$,
in the phase space,
there are two fixed points, corresponding to $\theta =\pi /2,$ $\phi =0$ and $\theta =\pi /2,$ $\phi =\pi $, respectively.
For weaker interaction $\Omega <2\kappa _{r}$,
two more stable fixed points appear which correspond to $\theta =\arcsin \sqrt{%
1-( \frac{\Omega }{2\kappa _{r}}) ^{2}},$ $\phi =0$ and $\theta
=\pi -\arcsin \sqrt{1-( \frac{\Omega }{2\kappa _{r}}) ^{2}},$ $%
\phi =0$, respectively.

\subsection{Dynamical stability}

In order to find out the fixed-point bifurcations,
we need to analyze the stabilities of the fixed points.
We first discuss the two fixed points ($\theta =\pi /2,$ $\phi =0,\pi $).
These points are interesting because they depend on neither
tunneling strength nor self-collision interaction strength.
We shall adopt the linear stability analysis that has wide applications in variable
nonlinear systems.
To begin with, we assume $p_{\phi }=p_{\phi
}^{0}+\delta _{p_{\phi }}$ and $\phi =\phi ^{0}+\delta _{\phi }$,
where $(p_{\phi }^{0}, \phi ^{0})$ denote one of the fixed points in the phase space,
$\delta_{p_{\phi }}$ and $\delta _{\phi }$ represent the deviations in the
population difference and relative phase from the fixed points, respectively.
With Eqs.\thinspace (\ref{dynamic1}) and (\ref{dynamic2}),
we obtain the linearized equation%
\begin{equation}
    \frac{\partial }{\partial t}\binom{\delta _{p_{\phi }}}{\delta _{\phi }}=M%
    \binom{\delta _{p_{\phi }}}{\delta _{\phi }}\text{,}
\end{equation}%
where $M$ is the Jacobian matrix%
\begin{equation}
    M=\left(
\begin{array}{cc}
    -\frac{\partial ^{2}H}{\partial p_{\phi }\partial \phi } & -\frac{\partial
    ^{2}H}{\partial \phi ^{2}} \\
    \frac{\partial ^{2}H}{\partial p_{\phi }^{2}} & \frac{\partial ^{2}H}{%
    \partial p_{\phi }\partial \phi }%
    \end{array}%
    \right) \text{.}  \label{Jacobian}
\end{equation}

The eigenvalues of the linearized equation may be real, pure or complex
imaginary. As is well known, the eigenvalues of the Jacobian matrix depict
the types and stabilities of the fixed points. By substituting
$\theta =\pi /2,$ $\phi =0,\pi $ into Eq.\thinspace (\ref%
{Jacobian}) and calculating the eigenvalues, we can find that the fixed
point $\theta =\pi /2,$ $\phi =0$ is stable for $\Omega >2\kappa _{r}$. When
$\Omega <2\kappa _{r}$, it becomes unstable. While for the fixed
point $\theta =\pi /2,$ $\phi =\pi $, it is always stable.
Company to the condition that two new fixed points emerge for $\Omega <2\kappa _{r}$,
the classical bifurcation condition then can be obtained as
\begin{equation}
    \Omega=2\kappa_{r}.
\end{equation}

\begin{figure}[tbp]
\includegraphics[width=8.5cm,clip]{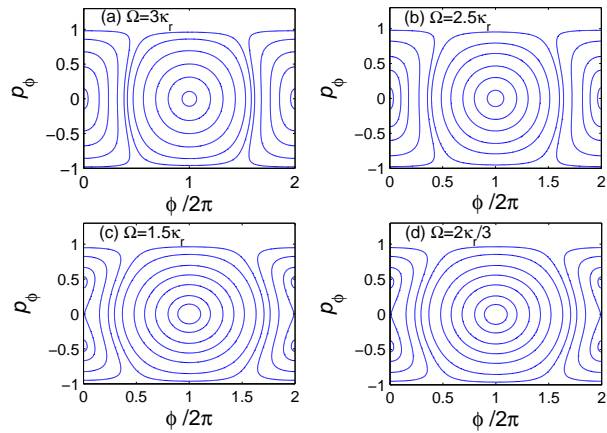}
\caption{(Color online) Trajectories of the Hamiltonian
system in the phase space for various parameters. $p_{\phi}$ corresponds to the double well
population difference and $\phi$ represents the phase difference between the two wells.}
\label{fig1}
\end{figure}

The stabilities of the fixed points can also be seen from the trajectories of the Hamiltonian.
In Fig.(\ref{fig1}), we plot the evolution trajectories
in the plane of $\theta $ and $\phi $ for various tunneling strengths.
It shows that a fixed point is stable if the evolution trajectories are loops around a fixed point;
otherwise it is unstable.
Note that all the stabilities of the fixed points which obtained by
the numerical simulation are consistent with the above theoretical analysis.

\subsection{Dynamical transition between JO and ST regimes}
Now we consider the transition from JO to ST regime.
In the JO regime, the population difference oscillates symmetrically between two wells,
and its average is zero.
In the ST regime, the average of the population difference is nonzero,
which can be obtained by the following condition~\cite%
{A. Smerzi}%
\begin{equation}
    \Omega \sin \theta _{0}\cos \phi _{0}+\kappa _{r}\cos ^{2}\theta _{0}>\Omega
\label{JO ST}
\end{equation}%
with $\theta _{0}$ and $\phi _{0}$ being the initial condition. From Eq. (\ref{JO ST}),
we get the critical point of the transition%
\begin{equation}
    \Omega _{c}=\frac{\kappa _{r}\cos ^{2}\theta _{0}}{1-\sin \theta _{0}\cos
    \phi _{0}}\text{.}  \label{MST_criterion}
\end{equation}%
For the initial value $\theta _{0}=0,$ $%
\phi _{0}=0$, the transition parameter is $\Omega_{c} =\kappa _{r}$ which is consistent
with the result of the model obtained by the mean field approximation~\cite{G. J. Milburn,L. B. Fu}%
. For $\theta _{0}=\pi /6,$ $\phi _{0}=0$, $\Omega_{c} =%
\frac{3\kappa _{r}}{2}$. If $\phi _{0}=\pi $ and $\theta _{0}=\pi/6$,
the corresponding critical value becomes $\Omega_{c} =\frac{\kappa _{r}}{2%
}$. It shows that the critical point can be adjusted by the relative phase between two wells,
and the relative phase can be experimentally adjusted by using a \textquotedblleft phase-imprinting\textquotedblright \ method \cite%
{Biao Wu}.

We emphasize that the particle number is large in the above classical analysis,
however, in the practical experiment, the number of particles is finite.
In the following discussions,
we consider the quantum dynamic of the QFI to investigate the dynamical stability of the fixed point and the
transition from JO to ST.
We will consider two cases that the initial SCSs are chosen as $\left \vert \theta =\pi
/2,\phi =0\right \rangle $ and $\left \vert \theta =0,\phi =0\right \rangle $%
. The SCS $\left \vert \theta =0,\phi =0\right \rangle $ is simply a Dicke
state $\left \vert j,-j\right \rangle $ in which all atoms lie in one
of the wells. $\left \vert \theta =\pi /2,\phi =0\right
\rangle $
is a state with a well-defined phase difference $\phi =0$ and
is a phase state of a two-mode boson system.
\begin{figure}[tbp]
\includegraphics[width=9cm,clip]{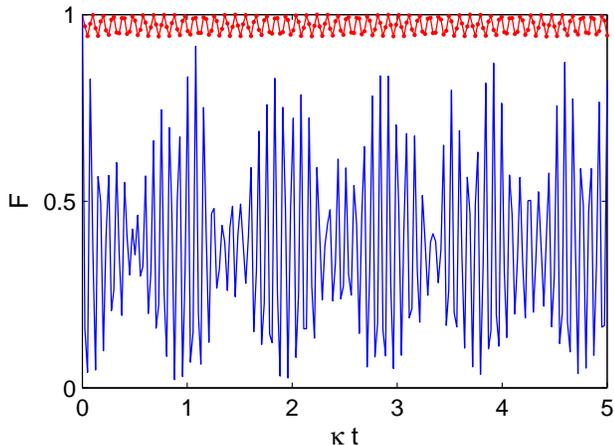}
\caption{(Color online) Fidelity of the overlap for different interaction of
the two-mode BEC. The blue curve represents a weaker interaction with $\Omega =%
\protect \kappa _{r}$, whereas the red curve with dots represents a stronger
interaction with $\Omega =4\protect \kappa _{r}$. For both cases the initial
state of the system is the SCS $\left \vert \protect \theta =\frac{\protect \pi }{2},\protect \phi %
=0\right \rangle $ and the number of atom is $N=100$.}
\label{fig2}
\end{figure}
\section{Effect of stability on Quantum Dynamic}

First we investigate the effect of stability on quantum dynamic of the system. We start
with the initial state as $\theta =\pi /2,$ $\phi =0,$ (i.e. $\left \vert j,j\right \rangle _{x}$) which corresponds to a fixed point.
Such a state can be
realized by applying a two photon $\pi /2$ pulse to the state $\left \vert
j,j\right \rangle $ with all the atoms in the internal state $\left \vert
F=1,m_{F}=1\right \rangle $ ~\cite{A. Soensen,U. V. Poulsen,Y. Li,A.
Widera,D. S. Hall}. For this initial state, in classical analogue,
it will change its stability at the classical bifurcation.

\subsection{Effect of stability on Fidelity}

In treating the quantum dynamical problem, it is helpful to bear in mind
some results from quantum information theory concerning fidelity. Fidelity
has been widely studied for the problems of transition probability in
quantum mechanics. Mathematically, for two pure states, the fidelity is
defined as
\begin{equation}
    F=\left \vert \left \langle \Psi \right \vert \left. \Phi \right \rangle
    \right \vert \text{.}
\end{equation}%
For the system with the initial wave function $\left \vert j,j\right \rangle
_{x}$, the state at arbitrary time $t$ can be expanded as \cite{Jin}%
\begin{equation}
    \left \vert \Psi (t)\right \rangle =\sum_{m}c_{m}(t)\left \vert j,m\right
    \rangle \text{,}
\end{equation}%
and the amplitudes $c_{m}(t)$ obeys%
\begin{eqnarray}
    i\dot{c}_{m}(t) &=&2\kappa m^{2}c_{m}(t)+\frac{\Omega }{2}c_{m-1}(t)\sqrt{%
    (j+m)(j-m+1)}  \notag \\
    &&+\frac{\Omega }{2}c_{m+1}(t)\sqrt{(j-m)(j+m+1)}\text{.}
\end{eqnarray}%
With the help of Eq.\thinspace (\ref{CSS}), the amplitude for the initial
SCS can be calculated as%
\begin{equation}
    c_{m}(0)=\frac{1}{2^{j}}\left( _{m+j}^{\text{ }2j}\right) ^{1/2}\text{.}
    \label{cm(0)}
\end{equation}%
According to the Heisenberg function, we get%
\begin{eqnarray}
    \dot{J}_{x} &=&-2\kappa (J_{y}J_{z}+J_{z}J_{y})\text{,} \\
    \dot{J}_{y} &=&2\kappa (J_{x}J_{z}+J_{z}J_{x})-2\Omega J_{z}, \\
    \dot{J}_{z} &=&\Omega J_{y}\text{.}
\end{eqnarray}

Unfortunately, for this model, it cannot be solved exactly for a many-particle case.
Numerical results of the wave function overlap between the
initial and the evolved state $F=\left \vert \left \langle \pi
/2,0\right \vert \left. \Psi (t)\right \rangle \right \vert $ are plotted in
Fig.\thinspace (\ref{fig2}).
The red curve with dots correspond to $\Omega=4\kappa _{r}$ and
blue one correspond to $\Omega =\kappa _{r}$.
For $\Omega=4\kappa _{r}$, it can be seen that the fidelity oscillates around the value $1$.
It indicates that the dynamic does not take the state far away from the initial state.
According to the stability analysis in Sec.\thinspace III, the fixed point which corresponds to the initial state
is stable in the regime $\Omega>2\kappa_{r}$.
When $\Omega =\kappa _{r}$, we can find that the fidelity oscillates between $0$ and $1$,
and the amplitude of the oscillations is inhomogeneous.
In the classical analogue, the fixed point is unstable in the regime $\Omega <2\kappa _{r}$.
From the above analysis,
we find that there is a perfect classical-quantum correspondence.

\begin{figure}[tbp]
\includegraphics[width=8.2cm,clip]{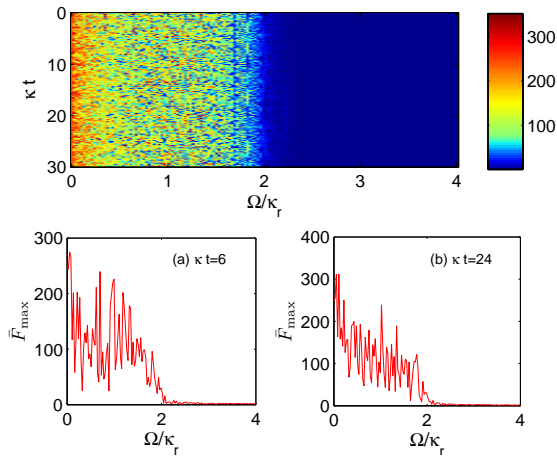}
\caption{(Color online) Top panel shows interaction-time plot of maximal
mean QFI $\bar{F}_{\max }$. The bottom shows the $\bar{F}_{\max }$ as a
function of $\Omega /\protect \kappa _{r}$ with rescaled time (a) $%
\protect \kappa t=6$ and (b) $\protect \kappa t=24$. Initial state of the
system is SCS $\left \vert
\protect \theta =\frac{\protect \pi }{2},\protect \phi =0\right \rangle $ and
the number of atom is $N=500$.}
\label{fig3}
\end{figure}

\subsection{Effect of stability on QFI}

Now, we consider the effect of stability on the QFI. Numerical results of the
maximal mean QFI $\bar{F}_{\max }$
as a function of $\frac{\Omega }{2\kappa _{r}}$ and $\kappa t$ are plotted in Fig.\thinspace (\ref{fig3}).
Obviously, the dynamic of the quantum system can be well illustrated by the maximal mean QFI.
As is shown in the top panel of Fig.\thinspace (\ref{fig3}),
the behaviors of the maximal mean QFI is quite different on the two sides
of the classical bifurcation point $\Omega =2\kappa _{r}$.
To clearly depict the phenomenon, at the bottom of Fig.\thinspace (\ref{fig3}),
we plot $\bar{F}_{\max }$ as a function of $%
\Omega /\kappa _{r}$ for two arbitrary rescaled time $\kappa t=6$ and $%
\kappa t=24$, respectively.
These two figures show that $\bar{F}_{\max }$ behaves irregular oscillations
in the regime $\Omega <2\kappa _{r}$,
in which the fixed point is unstable.
When $\Omega>2\kappa _{r}$, the fixed point is stable and we can find $\bar{F}_{\max }$
oscillates with time around the initial value.
Moreover, we can find that the value of the maximal mean QFI in the stable regime
is much smaller than that in the unstable regime.
The phenomenon can be understand by the fidelity which have been discussed in the above.
In the stable regime, the evolved state is close to the initial SCS,
and the $\bar{F}_{\max }$ is small.
While in the unstable regime, the evolved state far away from the initial SCS,
and the $\bar{F}_{\max }$ become bigger.
These results suggest that the QFI can be well used to characterize
the stability of the fixed point in this model.
\begin{figure}[tbp]
\includegraphics[width=8.5cm,clip]{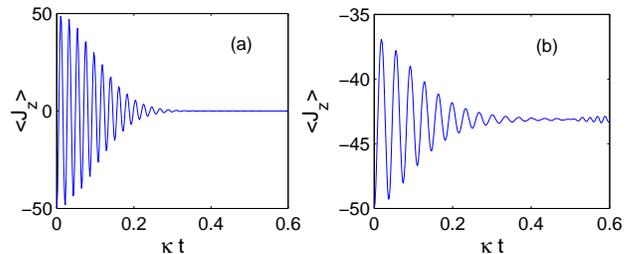}
\caption{(Color online) Quantum mechanical time evolution of $\left \langle
J_{z}\right \rangle $ in the ST and JO regimes for the initial condition $%
\left \vert j,-j\right \rangle $, corresponding to all atoms initially in
one of the wells. The plots in
terms of a dimensional parameter $\protect \kappa t,$ with (a) $\Omega =3%
\protect \kappa_{r} ;$ (b) $\Omega =2\protect \kappa_{r} /3$. The number of atom is $N=100$.}
\label{fig4}
\end{figure}
\section{Dynamical transition between JO and ST regimes}

\subsection{population difference in different regimes}

Below, we consider another dynamical problem, the transition from JO to ST regime.
Firstly, we investigate the population difference between two wells.
We choose $\left \vert \theta=0 ,\phi=0 \right \rangle$ as the initial state,
which corresponds to all atoms localized in one well.
In this case, it is clearly that for $\Omega =3\kappa
_{r}$ the initial state is related to the JO regime, whereas for $\Omega
=2\kappa _{r}/3$ it is related to the ST regime.
In Fig.\thinspace (\ref{fig4}a) and Fig.\thinspace (\ref{fig4}b),
we plot the quantum evolution of $\left \langle J_{z}\right \rangle $
as a function of $\kappa t$ for the JO and the ST regimes, respectively.
In the JO regime, $\left \langle J_{z}\right \rangle $
oscillates around zero during the evolution.
It indicates that there is no preferential tunneling to any of the wells.
While in the ST regime, $\left \langle J_{z}\right \rangle $ oscillates
around a non-zero value and only in the half plane.
It shows that part of the condensate is trapped in one of the wells.
From the above results, we find that the dynamical properties of
such a quantum system are quite different for the JO and the ST regimes.
\begin{figure}[tbp]
\includegraphics[width=8.2cm,clip]{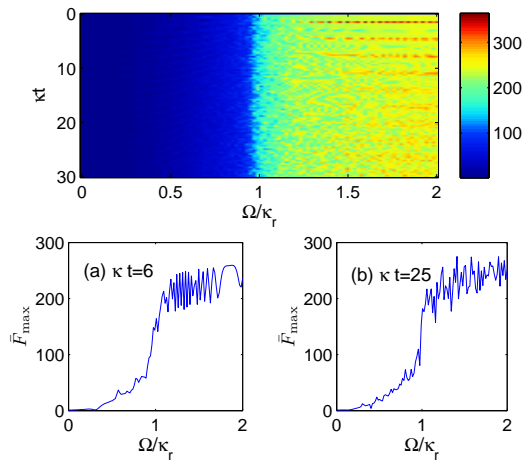}
\caption{(Color online) Top panel shows interaction-time plot of maximal
mean QFI $\bar{F}_{\max }$. The bottom shows the $\bar{F}_{\max }$ as a
function of $\Omega /\protect \kappa_{r} $ with rescaled time (a) $\protect%
\kappa t=6$ and (b) $\protect \kappa t=25$. Initial state of the system is
the SCS $\left \vert \protect \theta =0,%
\protect \phi =0\right \rangle $ and the number of atom is $N=500$. }
\label{fig5}
\end{figure}
\subsection{QFI in different regimes}

To well understand the quantum dynamical transition from JO to ST phenomenon,
we calculate the time evolution of the maximal mean QFI.
For a pure Dicke state $\left \vert j,m\right \rangle $,
the maximal QFI is obtained as%
\begin{equation}
    F_{\max }=2(j^{2}+j-m^{2}).
\end{equation}%
Then for the initial state $\left \vert j,-j\right \rangle $, the maximal mean QFI can be got as $\bar{F}_{\max }=1$.
Numerical results of the maximal mean QFI $\bar{F}_{\max }$ with different interactions are plotted in Fig.\thinspace (\ref{fig5}).
From the top of Fig.\thinspace (\ref{fig5}),
it is clearly seen that
the behavior of the maximal mean QFI are quite different in the JO and the ST regimes.
To well describe the behaviours of the maximal mean QFI for different regimes,
we plot the maximal mean QFI as a function of $\Omega/\kappa_{r}$ with two rescaled
time $\kappa t=6$ and $\kappa t=25$ in the bottom of Fig.\thinspace (\ref{fig5}).
From these two figures,
we can see that, for $\Omega <\kappa _{r}$, the maximal mean QFI increases slowly.
As the interaction strength approach the transition point $\Omega _{c}=\kappa _{r}$,
the $\bar{F}_{\max }$ increases quickly.
When $\Omega >\kappa _{r}$, it oscillates quickly and its value is
much larger than that in JO regime.
It is shown that the QFI can be served as an indicator of the transition from JO to ST regime.
Such abrupt change of the QFI also can be found in the phase transition of spin systems .

In experiments, so far, many observers have found number squeezing in this
model~\cite{G.-B. Jo}. Moreover, some researchers have shown that the number
fluctuation has a nontrivial relation with the phase fluctuation,
even for a single-mode light field~\cite{D. T. Pegg}. We hope that the transition of the maximal mean QFI
for this system can be observed in experiment in the future.

\section{Conclusion}

In conclusion, we have studied the dynamical evolution of a two mode
BEC in a symmetric double well. All the relevant fixed
points and their stabilities were analyzed. We investigated the dynamical behavior of
maximal mean QFI in this system which well described the stability of the fixed point. For the
initial state $\left \vert \theta =\pi /2,\phi =0\right \rangle $, the
numerical result showed that $\bar{F}%
_{\max }$ has a small oscillation around the initial value in the
$\Omega >2\kappa _{r}$ regime, while in the regime $\Omega <2\kappa
_{r}$, we found that $\bar{F}_{\max }$ displays a strong irregular
oscillation.

We also investigated the maximal mean QFI in the JO and the ST regimes which
corresponds to the different distribution of particles in two wells. For the
initial state $\left \vert \theta =0,\phi =0\right \rangle $, we found the
mean maximal QFI increases quickly at the critical point which corresponds to the boundary between the JO and the ST regimes.
From these results,
it showed that the QFI not only characterizes
the stability of the fixed point, but can also signal the presence the transition from
JO to ST regime.

\section{Acknowledgments}

X. Wang acknowledges support from the NFRPC with Grant No. 2012CB921602 and
NSFC with grant No. 11025527 and 10935010.

\end{document}